\documentclass[aps,prapplied,reprint,superscriptaddress]{revtex4-2}

\usepackage[cp1250]{inputenc}
\usepackage{bm,graphicx,color,ulem}
\usepackage{multirow}
\usepackage{upgreek}
\usepackage{amsmath}

\newcommand{\dlt}[1]{}
\newcommand{\del}[1]{}

\begin{document}


\title{Spatio-spectral metrics in electron energy loss spectroscopy as a tool to resolve nearly degenerate plasmon modes in dimer plasmonic antennas}

\author{Michal Hor\'ak}
\email{michal.horak2@ceitec.vutbr.cz}
\author{Andrea Kone\v{c}n\'a}
\affiliation{Central European Institute of Technology, Brno University of Technology, Purky\v{n}ova 123, 612 00 Brno, Czech Republic}
\author{Tom\'a\v{s} \v{S}ikola}
\author{Vlastimil K\v{r}\'apek}
\email{vlastimil.krapek@ceitec.vutbr.cz}
\affiliation{Central European Institute of Technology, Brno University of Technology, Purky\v{n}ova 123, 612 00 Brno, Czech Republic}
\affiliation{Institute of Physical Engineering, Brno University of Technology, Technick\'a 2, 616 69 Brno, Czech Republic}

\date{\today}

\begin{abstract}
Electron energy loss spectroscopy (EELS) is often utilized to characterize localized surface plasmon modes supported by plasmonic antennas. However, the spectral resolution of this technique is only mediocre, and it can be rather difficult to resolve modes close in the energy, such as coupled modes of dimer antennas. Here we address this issue for a case study of the dimer plasmonic antenna composed of two gold discs. We analyze four nearly degenerate coupled plasmon modes of the dimer: longitudinal and transverse bonding and antibonding dipole modes. With a traditional approach, which takes into account the spectral response of the antennas recorded at specific points, the modes cannot be experimentally identified with EELS. Therefore, we employ the spectral and spatial sensitivity of EELS simultaneously. We propose several metrics that can be utilized to resolve the modes. First, we utilize electrodynamic simulations to verify that the metrics indeed represent the spectral positions of the plasmon modes. Next, we apply the metrics to experimental data, demonstrating their ability to resolve three of the above-mentioned modes (with transverse bonding and antibonding modes still unresolved), identify them unequivocally, and determine their energies. In this respect, the spatio-spectral metrics increase the information extracted from electron energy loss spectroscopy applied to plasmonic antennas.
\end{abstract}


\maketitle

\section{Introduction}

Localized surface plasmons (LSP) emerge in metallic nanostructures, also known as plasmonic antennas (PAs), due to the coupling of surface charge in the metal and the related electromagnetic wave. They are manifested as broad spectral features in the optical response of PAs, observed e.g. by reflection and transmission spectroscopy, dark-field microscopy, scanning near-field optical microscopy, or electron beam spectroscopy. Scanning transmission electron microscopy (STEM) combined with electron energy loss spectroscopy (EELS) is an experimental technique allowing to characterize LSP in both the spatial and spectral domains.~\cite{doi:10.1146/annurev-physchem-040214-121612} Its spatial resolution (better than a nanometer) is excellent compared to the typical dimensions of PAs (tens to hundreds of nanometers) and decay lengths of LSP fields (hundreds of nanometers). However, the spectral resolution (around 100~meV, or around 10~meV for state-of-the-art instrumentation~\cite{KRIVANEK201960}) is rather low compared to linewidths of typical LSP resonances in the visible spectral range (around 100~meV).

In EELS, a monochromatized electron beam is transmitted through the sample. With a certain probability, an electron excites a LSP, losing characteristic energy. The transmitted electrons are subsequently characterized with a spectrometer and the loss probability is evaluated for each energy within the spectral region of interest. When the electron beam is scanned over the sample, the loss probability forms a three-dimensional EELS data cube (also referred to as the EELS spectrum image) with two spatial and one spectral dimensions. We note that it might be necessary to separate the contribution of LSP from the total spectrum, including bulk material losses and (nearly) elastically scattered electrons contributing to the zero-loss peak. When referring to the loss probability in the following, we will consider only the LSP contribution.

The loss probability $\Gamma(\omega)$ is directly related to the electric component of the LSP field at the frequency $\omega$ projected to the trajectory of the electron. More specifically,~\cite{RevModPhys.82.209}
\begin{equation}
\Gamma(\omega)=\frac{e}{\pi\hbar\omega}\int \mathrm{d}t\,\mathrm{Re}
\left \{ \exp(-\mathrm{i}\omega t) \mathbf{v} \cdot \mathbf{E}^\mathrm{ind}[\mathbf{r}_\mathrm{e}(t),\omega] \right \}
\end{equation}
where $\mathbf{E}^\mathrm{ind}[\mathbf{r}_\mathrm{e}(t),\omega]$ is the field induced by the electron moving with the velocity $\mathbf{v}$ at the position of the electron $\mathbf{r}_\mathrm{e}(t)$. For the electron with the trajectory perpendicular to the sample (along the axis $z$) it is an out-of-plane component of the field ($E_z$) that is relevant for the interaction. In such case, the electron trajectory is fully described by its $x$ and $y$ coordinates and the loss probability can be expressed as a three-dimensinal data cube $\Gamma(x, y, E)$, where the frequency of the field is replaced with the trasferred energy $E=\hbar \omega$ for convenience.

The loss probability has a close, although not always straightforward, relation to other quantities of interest, such as the (projected) photonic density of states~\cite{PhysRevLett.100.106804,PhysRevLett.103.106801} or extinction cross-section.~\cite{C3CS60478K} For the quantitative reconstruction of the electric field of LSP modes, electron beam tomography utilizes an electron beam rotated over the sample to reconstruct the spatial distribution of the electric field.~\cite{Horl2017,doi:10.1021/acs.nanolett.7b02979,doi:10.1021/acsphotonics.8b00125} 
Finally, with help of Babinet's principle it is possible to probe the magnetic component of the LSP field.~\cite{Horak2018SciRep,krapek_independent}

The most significant drawback of EELS is constituted by its mediocre spectral resolution, which makes it difficult to resolve spectrally overlapping LSP modes whose central energies are close to each other. A prototypical system supporting such overlapping LSP modes is a weakly coupled dimer PA consisting of two identical particles. The complexity of the loss spectrum is further increased when the components of the dimer themselves support overlapping modes, such is the case of disc-shaped particles. The lowest-order dipole mode in disc-shaped PAs is double degenerate. A nice illustration of this degeneracy is provided by the splitting of the mode observed when the symmetry of the disc is lowered e.g. by its gradual morphing into a triangle~\cite{doi:10.1021/nl502027r} or a crescent.~\cite{Krapek:15} Upon the formation of the dimer, the two pairs of dipole modes hybridize into a new set of modes: longitudinal dipole bonding (LDB), longitudinal dipole antibonding (LDA), transverse dipole bonding dipole (TDB), and transverse dipole antibonding (TDA) modes.~\cite{doi:10.1126/science.1089171,krapek_independent} The current oscillation for these modes is schematically shown in Fig.~\ref{figNO1}(c,d,e). Due to the interaction between the dipoles in the individual discs, the energies of the hybridized modes differ. Since the interaction between the longitudinal modes is stronger than between the transverse modes, a typical energy-ordering of the modes starting from the lowest energy one is LDB, TDB, TDA, and LDA.

Dimer PAs have been thoroughly studied both due to fundamental interest~\cite{doi:10.1126/science.1089171,krapek_independent,doi:10.1021/nl104410t,doi:10.1021/acsnano.5b02087,ZOHAR201426,doi:10.1021/jacs.0c13377,Song2021} and due to their intriguing applications in sensing,~\cite{ZOHAR201426} enhanced optical response,~\cite{Kinkhabwala2009} or strong light-matter coupling.~\cite{Bitton2020} From our perspective, it is worth revisiting the ability of EELS to resolve the dipole modes in a dimer PA. Song {\it et al.}~\cite{Song2021} reported a detailed study systematically varying the gap between two gold discs and demonstrated the experimental resolving of the LDB and LDA modes only for strongly coupled discs with the gap of 10~nm or less. Koh {\it et al.}~\cite{doi:10.1021/nl104410t} reported three nearly degenerate modes in a bowtie PA (a dimer of triangular prisms) with the gap between the prisms estimated to be about 25~nm. In this case, the modes are not experimentally resolvable, their identification is based on the simulations and even then it is rather illustrative. We note that their terminology for the modes differs from ours, with their dipolar bright, quadrupolar dark, and dipolar dark modes corresponding to ours LDB, a mixture of LD and TD, and LDA modes, respectively. Bitton {\it et al.}~\cite{Bitton2020} reported a silver bowtie PA with the gap of around 20~nm, and the LDB and LDA modes differing in energy by 0.3~eV. In this case, the modes were rather resolvable experimentally (also due to TD modes shifted to higher energies due to the low vertex angle of the bowtie) but their identification still required simulations. This overview, though not exhaustive, clearly demonstrates the lack of a methodology allowing to identify nearly degenerate LSP modes and to determine their energy solely from the experimental data.

In this manuscript, we address this issue by proposing several EELS metrics spanning both spatial and spectral degrees of freedom of a three-dimensional EELS data cube (containing the loss probability as a function of two spatial coordinates in the sample plane and one spectral coordinate). In a case study of a disc-shaped dimer PA, we test the performance of the metrics in the identification of the dipole modes. We demonstrate that the metrics allow to resolve and identify the LDB, TD (representing unresolved TDB and TDA), and LDA modes purely from the experimental data even for the gap size of 30~nm.

\section{Methods}

\subsection*{Fabrication of plasmonic antenna}

A gold disc-shaped dimer PA was fabricated on the substrate consisting of a silicon nitride membrane with the thickness of 30 nm and the lateral dimensions of $250 \times 250\ \mathrm{\upmu m^{2}}$. First, a gold layer with the thickness of 30 nm has been deposited by magnetron sputtering. The employed growth protocol produces a large-grain polycrystalline gold layer with optical properties comparable to monocrystalline gold in terms of Q factors of localized plasmon resonances.~\cite{Kejik:20} The dimer was fabricated by focused ion beam milling (using $\mathrm{Ga^+}$ ions at 30~keV) of the gold film in a dual beam system FEI Helios. The diameter of individual discs has been set to 275~nm and the edge-to-edge distance of the discs in the dimer has been set to 30~nm. The distance of the dimers from the boundary of the metal-free square has been at least 500~nm, which is a sufficient separation to prevent the interaction between the dimers or between the dimer and the surrounding metallic frame.~\cite{Krapek:15} Annular dark-field STEM image of the dimer is shown in Fig.~\ref{figNO4}(a).

Prior to the EELS measurements, the sample was cleaned in an argon-oxygen plasma cleaner for 20~seconds to prevent carbon contamination.~\cite{Horak2018}

\subsection*{Electron energy loss spectroscopy} 

EELS measurements were carried out in a scanning transmission electron microscope FEI Titan equipped with a GIF Quantum spectrometer. The microscope was operated at 120~kV in the scanning monochromated mode with the convergence semi-angle set to 10~mrad and the collection semi-angle set to 11.4~mrad. These settings are optimized for EELS characterization of PAs.~\cite{HORAK2020113044} The dispersion of the spectrometer was set to 0.01~eV per channel and the full-width at half-maximum of the zero-loss peak was found in the range from 0.10 to 0.12~eV. The probe current was adjusted to around 100~pA. The acquisition time of every spectrum was set to 0.5~ms to use the full intensity range of the CCD camera in the spectrometer and avoid its overexposure. The spatial resolution of the EELS data cube is determined by the pixel size, which was set to 5~nm. Such settings allowed the acquisition of one EELS data cube with a stable electron beam in a reasonable time.

The raw data containing electron counts recorded by the CCD shall be divided by the total number of electrons impinging the sample to obtain the loss probability, taking into account also the sensitivity of the detector. Since this approach is impractical, we utilize instead the electron counts of the zero-loss peak (with the energy integration window from $-1$~eV to +1~eV) to represent the total number of electrons. The division is then performed pixel-wise, i.e., independently for each position of the electron beam. In this way, we obtain the quantity proportional to the loss probability per channel, which is further divided by the energy interval of the channel (0.01~eV) to obtain the usual loss probability per electronvolt.

Further processing depends on the utilization of the EELS data cube. For EEL maps [Fig.~\ref{figNO3}] and metrics [Figs.~\ref{figNO5} and \ref{figNO6}] we integrate the data over the energy window of 0.1~eV around the target energy to suppress the noise. The zero-loss peak and background are not subtracted. They are both assumed to be constant over the mapped area, the zero-loss peak due to the normalization, and the background due to the homogeneity of the membrane. The EEL spectra [Fig.~\ref{figNO2}] are integrated over tens of pixels to reduce the noise. Next, experimentally determined background (including zero-loss peak) is subtracted to obtain a pure contribution of LSP. The background is measured for the electron beam impinging a bare membrane far from any plasmonic antennas.

\subsection*{Electromagnetic simulations}

Electron energy loss spectra have been calculated with the boundary element method (BEM) using a software package MNPBEM.~\cite{HOHENESTER2012370,WAXENEGGER2015138} In all simulations, the dimers are represented by two gold discs of the height of 30~nm on top of a 30-nm-thick silicon nitride membrane. The diameters of the discs (275~nm) and their edge-to-edge distance (30~nm) were matching the values set in the fabrication process. The dielectric function of gold was taken from Ref.~\cite{PhysRevB.6.4370} and the dielectric constant of the silicon nitride membrane was set equal to 4, which is a standard approximation in the considered spectral region.~\cite{doi:10.1021/nl502027r}

\section{Results and discussion}
\begin{figure}[ht!]
  \begin{center}
    \includegraphics[clip,width=\linewidth]{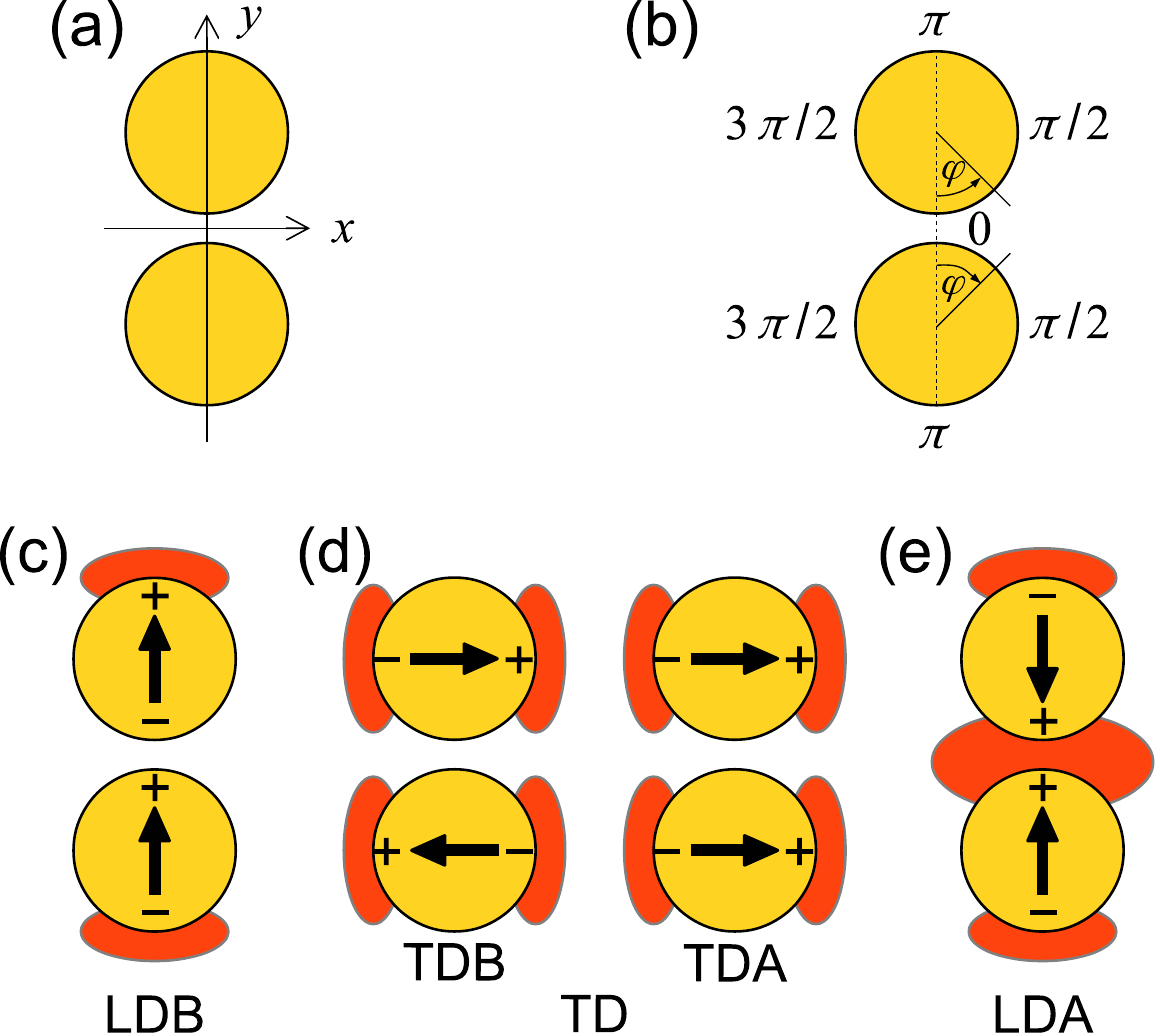}
	  \caption{\label{figNO1} (a,b) Cartesian (a) and angular (b) coordinates utilized in the definition of spatio-spectral metrics. (c,d,e) Schematic representation of (c) longitudinal dipole bonding (LDB), (d) transverse dipole (TD), and (e) longitudinal dipole antibonding (LDA) modes. Arrows represent charge oscillations, $+$ and $-$ signs represent the areas of charge accumulation, and red patches represent areas of pronounced loss probability for the respective modes. 
}
  \end{center}
\end{figure}

To facilitate the discussion of the spatial distribution of the loss probability we introduce a cartesian coordinate system with the axes $x$ and $y$ parallel with the transverse and longitudinal directions of the disc-shaped dimer, respectively, and the origin located in the middle of the gap between both discs. Further, we utilize two polar coordinate systems originating in the centers of the discs, with the zero angle corresponding to the center of the dimer, and increasing angle toward the right side of the discs (clockwise for the bottom disc and counter-clockwise for the top disc). All coordinates are represented in Fig.~\ref{figNO1}(a,b).

Fig.~\ref{figNO1}(c,d,e) shows intuitive schemes of the LDB, TD, and LDA modes. The modes are represented by two oscillating dipoles, one in each disc, oriented along the longitudinal ($y$) or transverse ($x$) direction. Current oscillates along the arrows, and charge accumulates near the boundaries of discs in the areas marked with $+$ and $-$ signs. According to Gauss law, this charge is a source of the electric field, which acts at the electron beam and results in the energy loss of the probing electrons. Consequently, the loss probability is large for the electron beam passing through the areas of accumulated charge. The only exception is the gap region of the LDB mode, where the opposite charges accumulated at the opposite sides of the gap between the discs effectively cancel each other due to the long-range character of Coulomb interaction with the probing electron, and the loss probability exhibits low values there. The regions of the large loss probability are displayed by red patches in Fig.~\ref{figNO1}(c,d,e). These regions are characteristic for each of the LDB, TD, and LDA modes (but not for the pair of TDB and TDA modes). The LDB mode features two loss-probability maxima along the longitudinal direction at the outer sides of the discs, the TD modes feature four maxima at the transverse sides of the discs, and the LDA mode exhibits three maxima along the longitudinal direction (two at the outer sides of the discs and one in the gap region between the inner sides).

\begin{figure}[ht!]
  \begin{center}
    \includegraphics[clip,width=\linewidth]{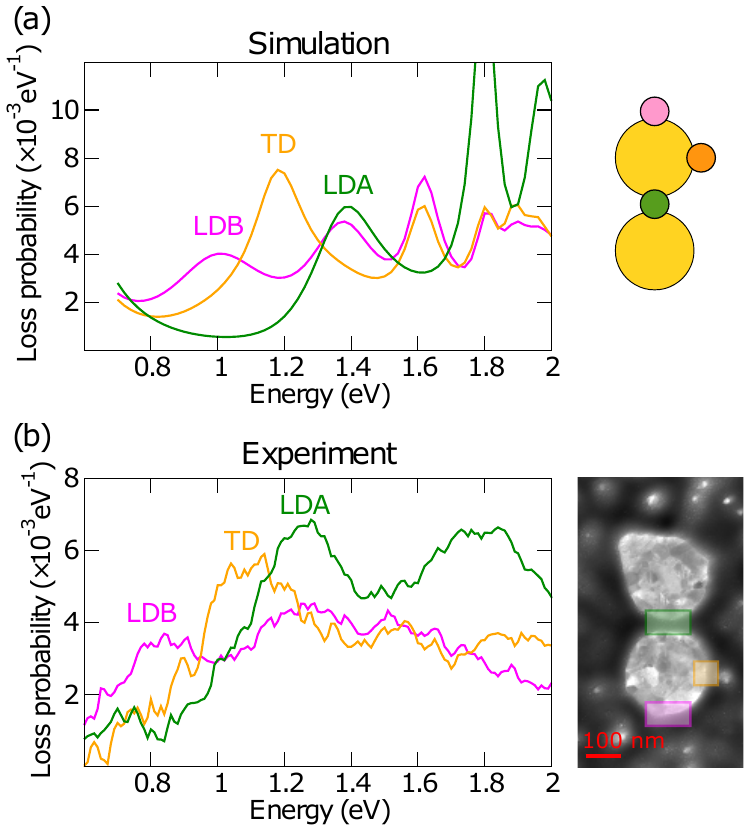}
	  \caption{\label{figNO2} Loss probability spectra of the disc-shaped dimer PA obtained from the numerical simulation (a) and the experiment (b). The insets show the electron beam positions for which the spectra were recorded. The color of the spot representing the electron beam position is identical to the color of the corresponding spectrum: pink for the outer longitudinal sides, orange for the transverse sides, and green for the gap.
}
  \end{center}
\end{figure}

\begin{figure*}[ht!]
  \begin{center}
    \includegraphics[clip,width=\linewidth]{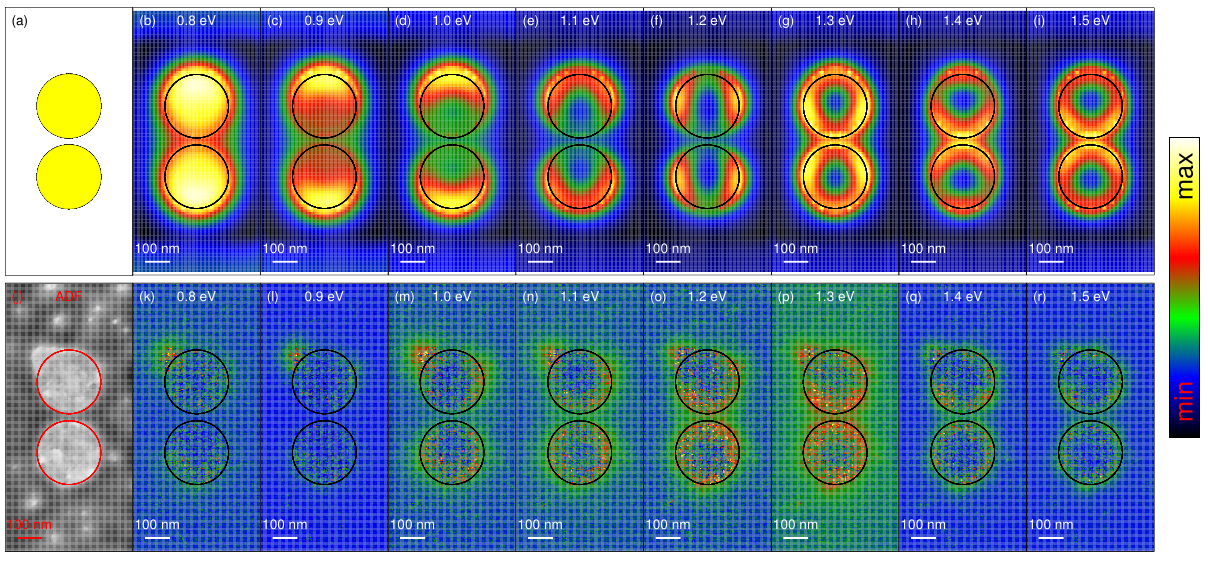}
	  \caption{\label{figNO3}
      Loss probability maps of the disc-shaped dimer PA obtained from the numerical simulation (top) and the experiment (bottom) for several loss energies.}
  \end{center}
\end{figure*}

Loss probability $\Gamma(x,y,E)$, a three-dimensional function of the electron beam position $x,y$ and the loss energy $E$, is typically visualized in the spectral domain as a spectrum for a specific beam position $x_0$, $y_0$,
\begin{equation}
\Gamma_\mathit{x_0,y_0}(E)=\Gamma(x=x_0,y=y_0,E),
\end{equation}
or in the spatial domain as a map for a specific loss energy $E_0$,
\begin{equation}
\Gamma_\mathit{E_0}(x,y)=\Gamma(x,y,E=E_0).
\end{equation}
The loss probability spectra obtained for our disc-shaped dimer and three distinct beam positions are shown in Fig.~\ref{figNO2}. To suppress the noise, the experimental spectra are integrated over multiple (several tens) pixels. The maps for several selected energies (later identified as the energies of LDB, TD, and LDA modes) are shown in Fig.~\ref{figNO3}. To suppress the noise, the experimental maps are integrated over multiple energy slices in the range 0.1~eV around the central energy (corresponding to 10 energy slices with our energy step of 0.01~eV. 

The theoretical loss spectra [Fig.~\ref{figNO2}(a)] exhibit three well separate peaks at the energy of 1.0~eV, 1.2~eV, and 1.4~eV. The peaks can be assigned to specific modes by a simple inspection of spatial maps. The peak at 1.0~eV with two longitudinal maxima [Fig.~\ref{figNO3}(c)] corresponds to the LDB mode, the peak at 1.2~eV with transverse maxima [Fig.~\ref{figNO3}(e)] corresponds to the TD modes, and the peak at 1.4~eV with two longitudinal maxima and one central maximum [Fig.~\ref{figNO3}(g)] corresponds to the LDA mode. The experimental spectrum [Fig.~\ref{figNO2}(b)] exhibits three quite resolvable features at the energy of 0.8~eV, 1.0~eV, and 1.2~eV. However, the related spatial maps do not clearly show features characteristic of considered modes and cannot be utilized for unambiguous peak-to-mode assignment. A natural question then arises whether any information on the modes is present and can be extracted from the experimental loss probability $\Gamma(x,y,E)$. We will show that the answer is positive when inspecting the loss probability in both spatial and spectral domains simultaneously.

\begin{figure}[ht!]
  \begin{center}
    \includegraphics[clip,width=\linewidth]{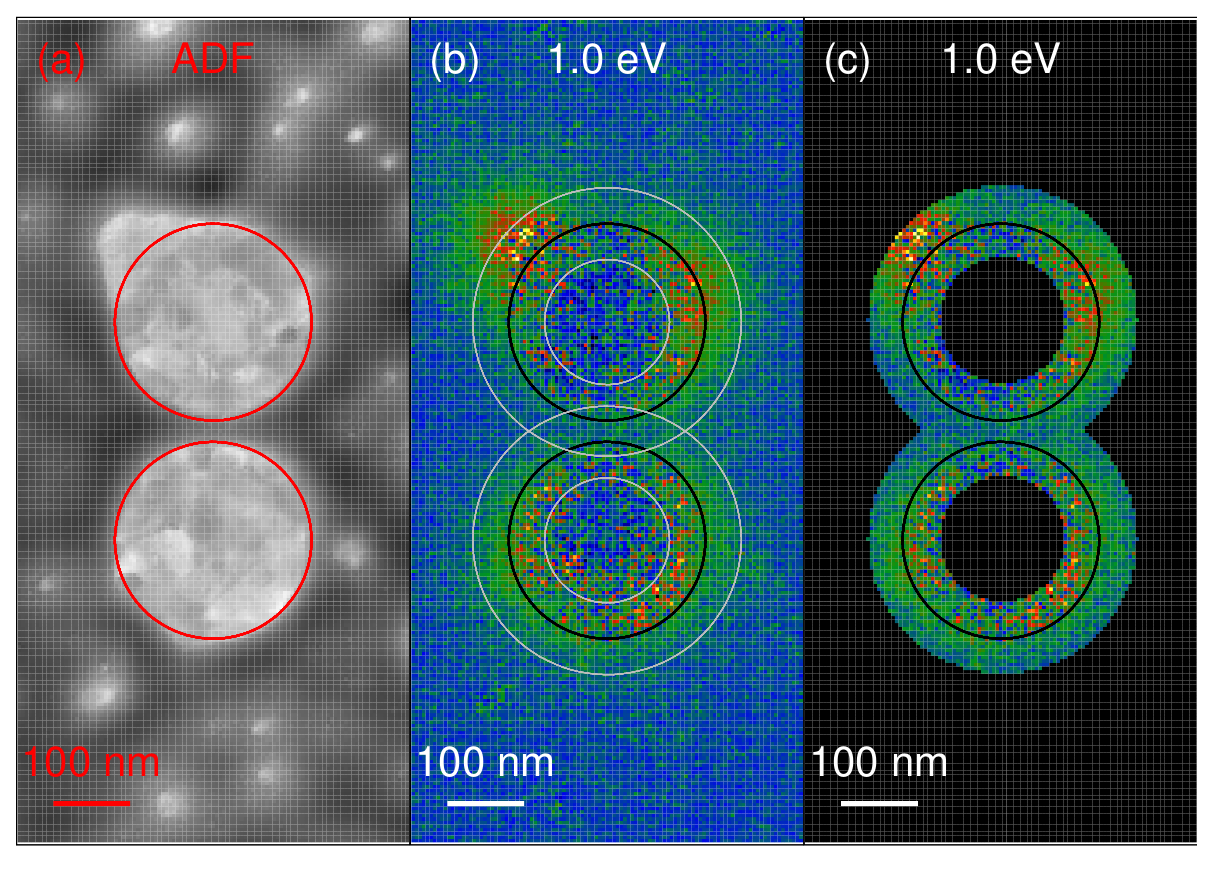}
	  \caption{\label{figNO4} (a) ADF STEM image of the disc-shaped dimer PA, with white/dark color representing gold/substrate. The red circles determine the boundaries of the discs utilized in the processing of the experimental loss probability. (b) A map of the loss probability at the energy of 1.0~eV. The black circles determine the boundaries of the discs and the gray circles mark the area with the high signal-to-noise ratio utilized for the calculation of the spatio-spectral metrics. (c) Same as panel b showing only the area with a high signal-to-noise ratio.}
  \end{center}
\end{figure}

One of the characteristic features of the considered modes is the spatial spread of the loss probability in the $x$ and $y$ (transverse and longitudinal, respectively) directions. This feature can be quantified by the so-called {\it absolute central moments}, similar to standard statistical moments, defined as follows. We first introduce the loss image weight $w(x,y,E)$, a quantity proportional to the loss probability but normalized to unity when integrated over a certain area of interest: $\int w(x,y,E) \,\mathrm{d}x\,\mathrm{d}y =1$. Clearly, 
\begin{equation}
w(x,y,E)=\frac{\Gamma(x,y,E)}{\int \Gamma(x,y,E)\,\mathrm{d}x\,\mathrm{d}y},
\end{equation}
where the integration goes over the area of interest. For simulations, the area of interest is a full simulation area. For experimental data, though, we selected two rings concentric with the plasmonic discs with a width of 100~nm where the signal-to-noise ratio is high [see Fig.~\ref{figNO4}(b,c)]. Next, we define the central coordinate of the loss image as
\begin{align}
x_\mathrm{C} &= \int x w(x,y,E)\,\mathrm{d}x\,\mathrm{d}y,\\
y_\mathrm{C} &= \int y w(x,y,E)\,\mathrm{d}x\,\mathrm{d}y, 
\end{align}
and the (energy-dependent) absolute central moments are defined as
\begin{align}
M_x(E) &= \int |x-x_\mathrm{C}| w(x,y,E)\,\mathrm{d}x\,\mathrm{d}y,\\
M_y(E) &= \int |y-y_\mathrm{C}| w(x,y,E)\,\mathrm{d}x\,\mathrm{d}y.
\end{align}

Before inspecting the actual values of the moments, we shall qualitatively discuss their expected properties. The longitudinal spread of the loss function shall be larger for the LDB mode than for the LDA mode due to the presence of a strong central maximum for the latter with near zero longitudinal (i.e., $y$) coordinate. Therefore, the longitudinal absolute central moment $M_y(E)$ shall be larger at the energy of the LDB mode compared to that at the energy of the LDA mode. Similarly, the longitudinal spread and the moment $M_y(E)$ of the TD modes shall be smaller than that of the LDB mode since the transverse maxima of the former are related with a lower longitudinal coordinate than the longitudinal maxima of the latter. We note that such a qualitative comparison is not possible for the TD and LDA modes. In the transverse direction, the TD mode shall feature a rather large transverse spread, accompanied by the large values of the transverse central moment $M_x(E)$, presumably larger than for the LDB and LDA modes. 

\begin{figure*}[ht!]
  \begin{center}
    \includegraphics[clip,width=\linewidth]{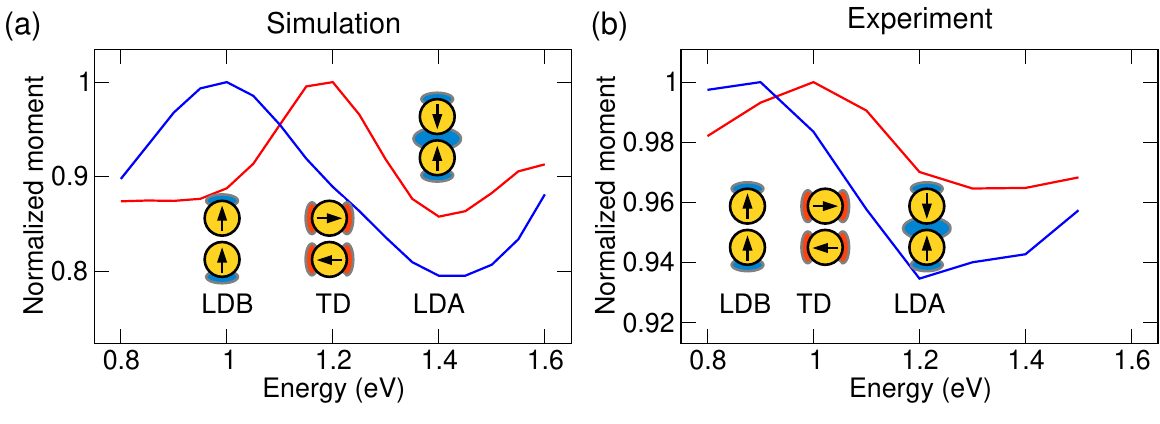}
	  \caption{Spectral dependence of the longitudinal absolute central moment $M_y(E)$ (blue) and the transverse absolute central moment $M_x(E)$ (red). The insets show the schemes of the LDB, TD, and LDA modes, and their position on the energy scale qualitatively corresponds to the actual mode energies. The colors of the patches representing the loss probability maxima correspond to their significance for the longitudinal (blue) and transverse (red) spectral features of the moments.\label{figNO5}}
  \end{center}
\end{figure*}

The spectral dependence of moments $M_x(E)$, $M_y(E)$ (normalized so that the maximum value in the inspected spectral region equals to unity) is shown in Fig.~\ref{figNO5}. The theoretical spectra are shown in the left panel. We will first discuss the longitudinal moment $M_y(E)$. As expected, its large value at the energy 1.0~eV corresponding to the LDB mode gradually decreases over the energy 1.2~eV corresponding to the TD modes to the energy 1.4~eV corresponding to the LDA mode. Strikingly, the maximum and minimum of the longitudinal moment corresponds exactly to the energies of the LDB and LDA modes, respectively. This suggests that the energies of the maximum and minimum can be used as the operational definition for the LDB and LDA modes energy if no more straightforward measure is available (such as in the case of the experiment). Similarly, the maximum of the transverse central moment at the energy of 1.2~eV agrees with the energy of the TD modes determined from the loss spectra (cmp. Fig.~\ref{figNO2} and related discussion) and can be used as the operational definition for the TD mode energy.

The experimental spectra of the moments correspond very well to the theoretical spectra, only with the reduced variance (3--5 times) over the examined spectral range, which is attributed to the experimental noise and/or background. This seemingly trivial statement represents an important finding. In contrast to theoretical maps of the loss probability, the experimental maps do not exhibit the apparent formation of the modes-identifying features such as the longitudinal, transverse, and central maxima. Still, the information allowing to identify the modes and to determine their energy can be extracted with judiciously defined metrics. The experimental energies of the LDB, TD, and LDA modes read 0.9~eV, 1.0~eV, and 1.2~eV, respectively (with a precision of 0.05~eV set by the half-width of the energy window used for the spectral integration of the experimental data), and correspond reasonably well to the positions of the peaks in Fig.~\ref{figNO2}(b) at 0.8~eV, 1.1~eV, and 1.3~eV. We suppose that the energies based on the absolute central moments are more relevant, since they take into account the spatial distribution of the loss probability, while the energies of the peaks in Fig.~\ref{figNO2} are more prone to the experimental noise and possible incomplete background subtraction.

The spatial distribution of the loss probability can be inspected also in polar coordinates. A non-trivial question arises how to set the origin of the polar coordinate system. The examined modes of the dimer are formed through the hybridization of the modes of individual discs. Therefore, it is meaningful to inspect the mode of a specific disc in its local coordinate system related to that disc. We use the following approach. We utilize the same area of interest as in the case of absolute central moments [i.e., two rings concentric with the plasmonic discs with a width of 100~nm; see Fig.~\ref{figNO4}(b,c))]. We also obtain the loss image weight as the loss intensity normalized to the unit integral over the area of interest. The loss image weight in each ring of the area of interest is processed separately in a local coordinate system of the relevant disc shown in Fig.~\ref{figNO1}(b) (the pixels from the intersection of the rings are processed twice). The loss image weight $w(x,y,E)$ is transformed to the (local) polar coordinates $r$, $\phi$,
\begin{equation}
w(x,y,E) \rightarrow w_\mathrm{U|L}(r,\phi,E),
\end{equation}
and then integrated over the radial coordinate, leaving only angular dependence,
\begin{equation}
W_\mathrm{U|L}(\phi,E)=\int w_\mathrm{U|L}(r,\phi,E)\,\mathrm{d}r.
\end{equation}
The indices $U$ and $L$ refer to the rings around the upper and lower disc, respectively, and the related local coordinate system. Finally, we define the zeros and positive directions of the angular coordinates in a way that respects the symmetry of the dimer. In this way, the angles $0$, $\pi/2$, $\pi$, and $3\pi/2$ correspond for both local coordinate systems to the center, right-hand (transverse) side, outer (longitudinal) side, and left-hand (transverse) side [see Fig.~\ref{figNO1}(b) and a related discussion]. This also allows us to inspect the total angular distributions of the loss image weight
\begin{equation}
W(\phi,E)=W_\mathrm{U}(\phi,E)+W_\mathrm{L}(\phi,E),
\end{equation}
instead of individual components. For brevity, we will refer to the quantity $W(\phi,E)$ as {\it angular loss weight} in the following. Now, the $W(\phi,E)$ shall feature maxima at the angle $\pi$ for the LDB mode, the angles $\pi/2$ and $3\pi/2$ for the TD modes, and the angles 0, $\pi$, and $2\pi$ for the LDA mode.

\begin{figure*}[ht!]
  \begin{center}
    \includegraphics[clip,width=\linewidth]{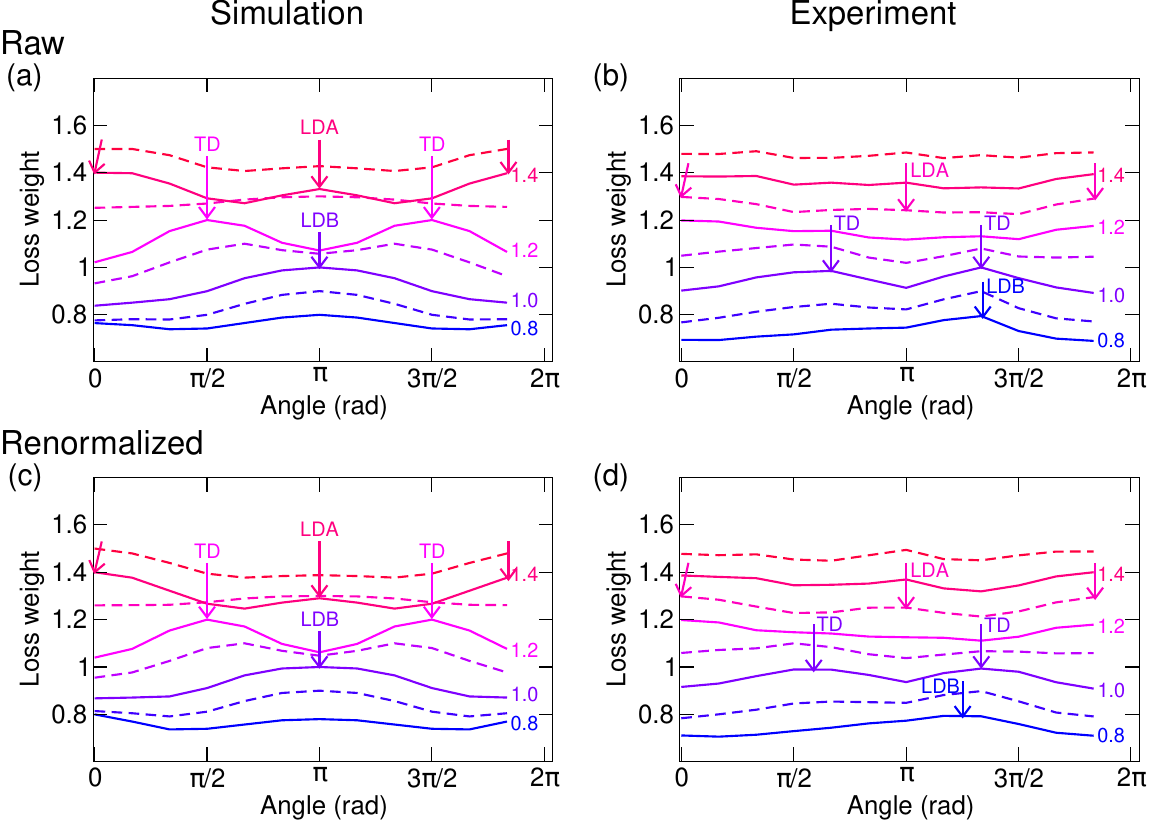}
	  \caption{(a,b) Angular loss weight obtained from the calculated (a) and experimental (b) data, taken at the energy indicated by the numbers (in eV units). The solid lines correspond to the energies of 0.8~eV, 1.0~eV, 1.2~eV, and 1.4~eV. The dashed lines correspond to the energies of 0.9~eV, 1.1~eV, 1.3~eV, and 1.5~eV. (c,d,) Renormalized angular loss weight obtained from the calculated (c) and experimental (d) data.\label{figNO6}}
  \end{center}
\end{figure*}

The angular loss weight calculated and experimentally retrieved for the dimer under study is shown in Fig.~\ref{figNO6}(a,b). The calculated angular loss weight evolves with the energy as expected, featuring a single peak at the angle of $\pi$ corresponding to the LDB mode at 0.9\,--\,1.0~eV, two peaks at the angles of $\pi/2$ and $3\pi/2$ corresponding to the TD modes at 1.2~eV, and two peaks at the angles $0$ (and also $2\pi$) and $\pi$ corresponding to the LDA mode at 1.4\,--\,1.5~eV. The energies correspond well to the energies determined from the spectra (Fig.~\ref{figNO2}) and absolute central moments (Fig.~\ref{figNO5}). The experimental weight angular distribution is qualitatively similar, with the signatures of the LDB, TD, and LDA modes at the energies of 0.8~eV, 1.0\,--\,1.1~eV, and 1.2\,--\,1.3~eV, respectively. However, the distribution profiles are distorted. Most prominently, they feature a strong peak around 4~rad for all energies, originating in the irregular shape of the dimer [a protrusion at the left upper corner of the dimer with a strong loss intensity observable in Fig.~\ref{figNO2}(a--h)]. To compensate for such irregularities, we proposed a {\it renormalized} total angular distribution of the loss weight
\begin{equation}
\tilde{W}(\phi,E)=W(\phi,E)/\sum_E W(\phi,E),
\end{equation}
i.e., we divide the {\it raw} weight angular distribution by the sum of the weight angular distributions over a range of energies. In this way, the renormalized weight $\tilde{W}(\phi,E)$ has a sum over the energy independent of the angle, and all angular variations are related purely to the properties of a specific mode. The renormalized total angular distribution of the loss weight is shown in Fig.~\ref{figNO6}(c,d). For calculated data we observe only insignificant differences between raw and renormalized weights, supporting the feasibility of the renormalization. For experimental data, the renormalized distributions are more symmetric and have peaks closer to the expected angles. The energies of the modes estimated from the renormalized weights are identical to those estimated from raw weights.

\begin{table}[ht!]
  \begin{center}
    \begin{tabular}{ccccccc}
      \hline
      \hline
         & \multicolumn{3}{c}{theory} & \multicolumn{3}{c}{experiment} \\
	 & LDB & TD & LDA & LDB & TD & LDA \\
      \hline
        spectrum & 1.0 & 1.2 & 1.4 & (0.8) & (1.1) & (1.3) \\
	ACS & 1.0 & 1.2 & 1.4 & 0.9 & 1.0 & 1.2 \\
	ALW & 0.9\,--\,1.0 & 1.2 & 1.4\,--\,1.5 & 0.8 & 1.0\,--\,1.1 & 1.2\,--\,1.3 \\
      \hline
      \hline
    \end{tabular} 
	  \caption{Mode energies in eV determined from the EEL spectrum and the spatio-spectral metrics: the absolute central moment (ACS) and the angular loss weight (ALW). The values in parentheses cannot be unequivocally assigned to the modes using only the experimental EEL spectrum.\label{tabNO1}}
  \end{center}
\end{table}

The spatio-spectral metrics introduced in the manuscript predict the mode energies in agreement with each other as well as in agreement with EEL spectra, as summarized in Table~\ref{tabNO1}. The metrics offer several benefits over the traditional EEL spectra. First, they are tailored to specific modes and describe their features in more detail than the spectra. Second, they utilize a larger part of the EELS data cube. In consequence, spatio-spectral metrics allow reliable identification of the nearly degenerate modes and a determination of their energy. With EEL spectra, this is possible using calculated data but not using noisy experimental data. Both metrics, the absolute central moment and the angular weight distribution, have their advantages and disadvantages. The absolute central moment is well suited for systems with naturally defined longitudinal and transverse directions, such as dimers or single rods. However, its predictive power will be limited for systems lacking such directions, e.g., triangular prisms or single spheres. On the other hand, the angular weight distribution is more general (as long as the natural central point of the system exists).

In addition to LSP modes, spatio-spectral metrics are applicable to other local excitations with non-trivial spectral and spatial dependence, such as the optical modes in dielectric nanoparticles~\cite{doi:10.1021/acsnano.1c06065} and localized vibrational modes supported by nanoparticles.~\cite{Lagos2017,doi:10.1126/science.abg0330} With some adaptation, the metrics can be applied also to data with additional dimensions, as in the case of spectral, spatial and angular dependence of localized surface plasmon modes characterized by cathodoluminescence.~\cite{10.1093/jmicro/dfw025,doi:10.1021/acs.nanolett.0c04084}

\section{Conclusion}

We have addressed an issue of resolving nearly degenerate localized surface plasmons using electron energy loss spectroscopy. More specifically, we have studied four plasmon dipole modes supported by a dimer of plasmonic discs. An inspection of the experimental loss probability individually in the spectral domain (i.e., loss spectra) and the spatial domain (i.e., loss maps) allowed neither unequivocal identification of the modes nor the reliable determination of their energies. Consequently, we have proposed two spatio-spectral metrics defined over both spatial and spectral domains of the loss probability. With their help, the identification of the modes and their energy was made possible. As a side benefit, the metrics require only the rudimentary processing of raw data with no need for background subtraction. We are convinced that simultaneous inspection of the loss probability in both spatial and spectral domains opens the way for a more detailed and reliable data analysis and thus significantly enhances the capability of electron energy loss spectroscopy.

\begin{acknowledgments}
We acknowledge the support by the Czech Science Foundation (grant No.~22-04859S), and by the Ministry of Education, Youth and Sports of the Czech Republic (project CzechNanoLab, No. LM2018110).
\end{acknowledgments}

\bibliography{manuscript}

\end{document}